\documentstyle[a4,12pt,psfig]{article}
\begin{document}

\newcommand{\al}{\alpha}
\newcommand{\bt}{\beta}
\newcommand{\1}{\vspace{1cm}\noindent}
\newcommand{\4}{\vspace{4cm}\noindent}
\newcommand{\be}{\begin{equation}}
\newcommand{\ee}{\end{equation}}
\newcommand{\ba}{\begin{eqnarray}}
\newcommand{\ea}{\end{eqnarray}}
\newcommand{\de}{\delta}
\newcommand{\dd}{\partial}
\newcommand{\ga}{\gamma}
\newcommand{\sg}{\sigma}
\newcommand{\db}{\bar{\partial}}
\newcommand{\fb}{\bar{f}}
\newcommand{\tet}{e_z^{~~a}}
\newcommand{\tec}{e_{\bar{z}}^{~~a}}
\newcommand{\teo}{e_0^{~~a}}
\newcommand{\ta}{\theta}
\newcommand{\ach}{\mbox{arccosh}}
\newcommand{\bz}{\bar{z}}
\newcommand{\ra}{\rightarrow}
\newcommand{\la}{\lambda}
\newcommand{\nn}{\nonumber}
\newcommand{\p}{\psi^\al(z)}
\newcommand{\pbar}{\bar{\psi}_\beta(z_o)}
\newcommand{\vs}{\vspace{0.2cm}}
\newcommand{\vsa}{\vspace{0.4cm}}
\newcommand{\vsb}{\vspace{0.8cm}}
\newcommand{\half}{\frac{1}{2}}
\newcommand{\eps}{\varepsilon}
\newcommand{\vfi}{\varphi}
\newcommand{\dr}{e^{~~a}_\mu}
\newcommand{\scon}{\omega^{~~a}_\mu}
\newcommand{\pijl}{\leftrightarrow}
\newcommand{\cd}{{\cal D}}
\newcommand{\cb}{{\cal B}}
\newcommand{\cc}{{\cal C}}
\newcommand{\veps}{\varepsilon}
\newcommand{\vx}{\vec{x}}
\newcommand{\vp}{\vec{p}}
\newcommand{\om}{\omega}
\newcommand{\ck}{{\cal K}^l_{m,(t)}(\xi,\vfi)}
\newcommand{\cdd}{\cd^l_{m,t}(H,\vfi,\xi)}
\newcommand{\vt}{\vartheta}
\newcommand{\Ml}{{\cal M}^\la_m(\eta,\vt)}
\newcommand{\Mr}{\overline{{\cal M}^{\la}_{m}}(\eta,\vt)}
\newcommand{\Mrs}{\overline{{\cal M}^{\la'}_{m'}}(\eta,\vt)}
\newcommand{\Mrr}{\overline{{\cal M}^{\la}_{m}}(\eta',\vt')}

\begin{titlepage}
\begin{flushright}
Preprint:THU-97/07\\
hep-th/9703058 \\
March 1997
\end{flushright}
\vsa
\begin{center}
{\large\bf Two Particle Quantummechanics in 2+1 Gravity \vs\\
           using Non Commuting Coordinates   \vsa\vsb\\}
           M.Welling\footnote{E-mail: welling@fys.ruu.nl\\
                     Work supported by the European Commission TMR programme
                                                         ERBFMRX-CT96-0045}
     \vsa\vsb\\
   {\it Instituut voor Theoretische Fysica\\
     Rijksuniversiteit Utrecht\\
     Princetonplein 5\\
     P.O.\ Box 80006\\
     3508 TA Utrecht\\
     The Netherlands}\vsb\vsa\\
\end{center}
\begin{abstract}
We find that the momentum conjugate to the relative distance between two
gravitating particles in their center of mass frame is a hyperbolic angle.
This fact strongly suggests that
momentum space should be taken to be a hyperboloid.
We investigate the effect of quantization on this curved momentum space.
The coordinates are represented by non commuting, Hermitian operators on this
hyperboloid. We also find that there is a smallest distance between the two
particles of one half times the Planck length.
\end{abstract}

\end{titlepage}

\section{Introduction}
The classical theory of 2+1 gravity was first considered in 1963 by
Staruszkiewicz \cite{Star}.
The subject was revived in 1984 by Deser, Jackiw and 't Hooft in their
cornerstone article \cite{DJH}. The basic feature that one can find in these
articles is that the gravitational field does not carry degrees of freedom. In
that sense it is topological theory. Pointparticle solutions can be constructed
by a `cut and paste' procedure. For instance for one particle at rest one
simply cuts out a wedgelike region from space time and identifies the
boundaries. More elaborate configurations are always characterized by elements
of the Poincar\'e group. An elegant procedure to construct multiparticle
solutions was put forward by 't Hooft \cite{Hooft0} who glued together flat
patches of Minkowski space using Lorentz transformations.

Achucarro and Townsend \cite{CS1} (1986) and Witten \cite{CS2} (1988) proposed
a Chern Simons gauge theory that is equivalent to 2+1 gravity.  Many of the
subsequent quantum models used this formulation and a great deal of progress
was made.
These models however concerned mainly with closed universes with non trivial
topology (handles) \cite{Carlip1}
and with the calculation of scattering amplitudes \cite{Hooft1, DesJac, VazWit,
Carlip2}. However no one passed the point of writing down a complete
Hilbert space for a multiparticle model. A consistent theory for the creation
and annihilation of particles was not considered at all to our knowledge.
However this is of course what we would ultimately like to understand. What is
the problem we are facing when trying to write down a Hilbert space for a two
particle system for instance? If we solve the Klein Gordon equation on a cone
for fixed energy we find a set of fractional Bessel functions times some
angular
function:
\be
\psi_E(r,\vt)=J_{\frac{n}{\al}}(kr)~e^{i\frac{n}{\al}\vt}
\ee
with $\hbar k=\sqrt{2ME}$ and $\al=1-\frac{E}{2\pi}$. However for different
energy the $\al$'s are different which implies that the wave functions cannot
be orthonormal. In this paper we will encounter a similar problem.

Another point that was frequently overlooked is the choice of canonically
conjugate variables. Because the Hamiltonian is not simply of the Klein Gordon
type, the conjugate momentum to for instance the distance of the particle from
the origin is not simply $Mv^a$ (mass times velocity). The first to take into
account the correct Hamiltonian and a pair of conjugate phase space variables
was 't Hooft in \cite{Hooft3}. He found that the momentum conjugate to the
distance of the particle from the origin is actually an angle! Using this he
quantized the particle on a spherical momentum space. The Hilbert space is then
simply spanned by the spherical harmomics $Y_{l,m}(\theta,\vfi)$.

The idea to use a curved momentum space goes back to 1947. In his paper Snyder
proposed de Sitter or anti de Sitter spaces for energy momentum space. He
introduced non commuting operators to represent the
coordinates $x,y,z,t$, turning configuration space partly into a lattice.
However the model was still covariant
with respect to the full Lorentzgroup! Recently Demichev considered this issue
in the light of quantumgroups
\cite{Dem}.

In this paper we will follow the same kind of reasoning as in \cite{Hooft3}. In
section one we calculate the conjugate momentum to the relative distance
between two particles and find a {\em hyperbolic} angle. This suggests
quantization on a hyperboloid. In section two we try to define Hermitian
operators that represent the coordinates $x,y$. The fact that momentum space is
curved (i.e. a hyperboloid) turnes these coordinates into non commuting
operators. We also define new momenta which are conjugate to $x,y$ only up
to order $\ell_P^2$ where $\ell_P$ is the Planck length. We also calculate the
commutator algebra between these operators. In the last section we compute a
complete set of basisfuntions that span the Hilbert space in the case when we
do
not take the boundary conditions into account properly, but only study the
effect of the curvature of momentum space. We find that the particles cannot
approach each other closer than $\frac{1}{2}\ell_P$. In the case when we do
take
into account the correct boundary conditions we find that we are stuck with a
similar problem as was the case with the fractional Besselfunctions: wave
functions with different energy are not orthogonal. Appendix A treats the case
of massless particles. In appendix B we derive a Dirac equation.

\section{Canonical conjugate coordinates}

In 1988 't Hooft showed in \cite{Hooft1} that we may describe two gravitating
particles in their center of mass (c.o.m.) frame by a relative distance vector
$\vec{r}$ which obeys a peculiar boundary condition. The effect of the
particles on the space time is that we should cut out a wedgelike region from
space time with a deficit angle $\bt=E$, where $E$ is the energy of the two
particles. The boundaries of the wedge should be identified in some way
depending on the value of the angular momentum $L$:
\be
\left( \begin{array}{c}
t'\\
x'\\
y' \end{array}\right)=
\left(\begin{array}{ccc}
1&0&0\\
0&\cos E&\sin E\\
0&-\sin E&\cos E\end{array}\right)
\left(\begin{array}{c}
t\\
x\\
y \end{array}\right)+
\left(\begin{array}{c}
L\\
0\\
0\end{array}\right)
\label{bc}
\ee
When we do quantummechanics in such a space we need to incorporate this
identification as a boundary condition on the wavefunction. Writing
\be
\psi=\sum_m \psi_m(r)e^{im\vt}
\ee
we find:
\be
\psi(r,\vt+2\pi)=e^{imE}\psi(r,\vt)\label{boundarycondition}
\ee
This leads to a revised quantization for the angular momentum:
\be
m=\frac{2\pi n}{2\pi-E}~~~~~n\in Z
\ee
He then proceeds to calculate the scattering amplitude using the Hamiltonian:
\be
H=\sqrt{p^2+M_1^2}+\sqrt{p^2+M^2_2}
\ee
Although the amplitude does not depend on the choice of this Hamiltonian it is
not the correct one and only valid in the low energy regime. In this paper we
want to investigate the two particle Hilbert space using a proper Hamiltonian.
If we define time at infinity, Carlip \cite{Carlip2} has shown that the
Hamiltonian is given by the total deficit angle at infinity\footnote{Waelbroeck
has shown however, \cite{Waelbroeck} using his lattice model for 2+1 gravity,
that different choices for time may lead to different Hamiltonians and
different
quantizations. For one choice time is quantized, for another it is not!}.
It is expressed in the following way:
\be
e^{iHJ_0}=e^{ip_1^aJ_a}e^{ip^a_2J_a}
\ee
The $J_a$ are the generators of the group SO(2,1) and
\be
p_i^a=(M_i\cosh\xi_i,M_i\sinh\xi_i\cos\vfi_i,M_i\sinh\xi_i\sin\vfi_i)
\ee
where $\xi$ is the rapidity of the particle. This can be rewritten as:
\ba
\cos(\frac{H}{2})&=&\cos\mu_1\cos\mu_2-\sin\mu_1\sin\mu_2
(\frac{p_1^a ~p_{2a}}{M_1 M_2})\label{a}\\
\sin(\frac{H}{2})\frac{P^a}{|P|}&=&\cos\mu_1\sin\mu_2\frac{p_1^a}{M_1}+
\cos\mu_2\sin\mu_1\frac{p^a_2}{M_2}+
\sin\mu_1\sin\mu_2\eps^{abc}\frac{p_{1b}~p_{2c}}{M_1 M_2}\label{b}
\ea
where we introduced $\mu_i\equiv\frac{M_i}{2}$. To simplify things a bit we
will now assume that both particles have the same mass: $\mu_1=\mu_2$.
In appendix A we will treat the massless case, i.e. $\mu_1=\mu_2=0$.
Furthermore, in the c.o.m. frame we have that $E_1=E_2$, but as we will show
shortly, not $\vec{p}_1=-\vec{p}_2$. Let us parametrize:
\ba
p_1&=&(E,p,0)\\
p_2&=&(E,p\cos\vfi,p\sin\vfi)
\ea
with $E=\sqrt{p^2+M^2}$. In the c.o.m. frame we also have: $P^i=0$. Using this
fact we can calculate the the angle $\vfi$:
\be
\cos\vfi=\frac{E^2\sin^2\mu -M^2\cos^2\mu }{E^2\sin^2\mu +M^2\cos^2\mu}
\label{hoek}
\ee
This is clearly an effect of the gravitational field, not present in the case
of two non gravitating particles. Equation (\ref{a}) can be written as:
\be
\cos(\frac{H}{2})=\cos^2\mu-\sin^2\mu(E^2-p^2\cos\vfi)
\ee
Using (\ref{hoek}) we find for this:
\be
\cos(\frac{H}{2})=\frac{\cos M-\sin^2\mu\sinh^2\xi}{1+\sin^2\mu\sinh^2\xi}
\label{energy}
\ee
One can check that:
\ba
\xi\ra 0&\Rightarrow& H\ra 2M\\
\xi\ra \infty &\Rightarrow & H\ra 2\pi
\ea
as is to be expected.
Now that we know the Hamiltonian as a function of $p$ (or $\xi$) we can
calculate the canonically conjugate momentum $q$ to the relative distance $r$.
The other coordinate we will use is an angle $\vt$ and the conjugate angular
momentum L. Because the particles are in free motion we know what the velocity
is:
\be
\frac{d}{d t}r=\sqrt{(\vec{v_1}^2-\vec{v_2}^2)^2}
\ee
with
\ba
\vec{v_1}&=&(\tanh\xi,0)\\
\vec{v_2}&=&(\tanh\xi\cos\vfi,\tanh\xi\sin\vfi)
\ea
We find:
\be
\frac{d}{d t}r=\frac{2\cos\mu~\tanh\xi}{\sqrt{1+\sin^2\mu~\sinh^2\xi}}
\ee
Next we solve:
\be
\frac{d}{d t}r=\frac{d H}{d q}\Rightarrow
q=\int d\xi \frac{-2}{\dot{r}\sin(\frac{H}{2})}\frac{d \cos(\frac{H}{2})}{d
\xi}
\ee
\be
=\int d\xi \frac{2\sin\mu~\cosh\xi}{\sqrt{1+\sin^2\mu~\sinh^2\xi}}=
2\cosh^{-1}[\sin\mu~\sinh\xi]
\ee
In other words:
\be
\sinh(\frac{q}{2})=\sin\mu~\sinh\xi
\ee
So we find that the momentum $q$ conjugate to the relative distance is a
hyperbolic
angle. Substituting this in the formula for the energy (\ref{energy}) we see:
\be
\cos(\frac{H}{4})=\frac{\cos\mu}{\cosh(\frac{q}{2})}\label{energy2}
\ee
If we set $q=2\eta$ and $H_P=\frac{1}{2}H$ this is precisely the mass shell
condition for one particle in the polygon approach to 2+1 gravity
\footnote{In the polygon approach the conjugate variables are $L$ and $2\eta$.
$L$ is the length of an edge of the wedge; $\eta$ is the rapidity with which
this edge moves perpendicular to $L$.}
\cite{Hooft0, Welling}.
We can therefore conclude that the configuration space for the relative
coordinate $r$ is a `spinning cone'. The value of the `spin' is the angular
momentum of the two particles, the excised angle equals the total energy
(see (\ref{bc})) and the mass shell condition is given by (\ref{energy2}).
We stress that the configuration space for this `effective c.o.m. particle'
is not space time surrounding the two physical particles. This is a
`double cone'.
The quantization of this effective particle in
the following
can therefore also be interpreted as an alternative program for the
quantization on a sphere as was done by 't Hooft \cite{Hooft2}.
It might seem confusing that he found that the momentum conjugate to the
distance of a particle as seen from a fixed origin is actually an angle instead
of a hyperbolic angle. The relative distance between two particles (or the edge
length $L$ in the case of one particle)  is however a different coordinate and
as such has a different conjugate momentum.
Another peculiar aspect of (\ref{energy2}) is that $H$ is an angle and ranges
from $2M$ to $2\pi$. This fact was used by 't Hooft in \cite{Hooft2} to show
that time is discrete. We will also adopt this point of view although it is not
essential in this paper.
Note finally that for small momentum and mass we can expand the `mass shell'
relation (\ref{energy2}):
\be
H=2\sqrt{q^2+M^2}~~~~~q,M<<1\label{KG}
\ee
which is of course the Klein Gordon equation, only valid in the low
energy regime.

\section{Curved momentum space and non commuting coordinates}

In the previous section we have seen that a convenient set of canonically
conjugate coordinates is given by:
\ba
q&\leftrightarrow& r\label{can}\\
\vt&\leftrightarrow& L
\ea
Looking at the mass shell equation we see that $\frac{q}{2}(\equiv \eta)$ only
appears as the argument of a sinehyperbolic. It is therefore wise to choose a
basis in the Hilbert space upon which $\sinh\eta$ has a natural action. In much
the same spirit as 't Hooft does in \cite{Hooft3} we will introduce a curved
momentum space on which we will perform the quantization and study its
consequences. The surface we choose as our momentum space is a hyperboloid
embedded in $R^3$ as follows:
\ba
Q_1&=&\sinh\eta\cos\vt\\
Q_2&=&\sinh\eta\sin\vt\\
Q_3&=&\cosh\eta
\ea
\be
Q_3^2-Q_1^2-Q_2^2=1\label{hyp}
\ee
Because we like to keep translational invariance in the timelike direction we
take as our energy momentum space ${\cal M}_{H,p}=H^2_1\times S^1$.
$H^2_1$ denotes a hyperboloid with two negative signs and one positive sign in
(\ref{hyp}). $S^1$ is a circel because the energy is an angle. To compare, 't
Hooft considered quantization on $S^2\times S^1$ and $S^3$. Actually
quantization on $H^2_2\sim SL(2,R)$ is also possible \cite{MatWel}. Because we
will do harmonic analysis on the hyperboloid it is usefull to remark that
$H^2_1\sim$SO(2,1)/SO(2). Fortunately there is a lot of literature on harmonic
analysis on groups and cosets of groups at our disposal. Remember that
the group space is momentum space and not configuration space as is usually
considered.

If we define, in addition to the $Q_i$ above, the following $Q_0$:
\be
Q_0=\cos\mu~\tan \tilde{H}
\ee
where $\tilde{H}\equiv\frac{H}{4}$, we find that the mass shell condition
becomes the Klein Gordon equation:
\be
Q_0^2-Q_1^2-Q_2^2=\sin^2\mu
\ee
which is of course invariant with respect to a Lorentz transformation of
$(Q_1,Q_2,Q_0)$
\footnote{This is a Lorentz transformation of the effective particle which is
not a Lorentz transformation of ordinary space time}.
In appendix B we will construct a linear Dirac equation.
The fact that this is indeed a Lorentzvector can be seen from
the following:
\ba
Q_1&=&\sinh\eta\cos\vt=\sin\mu\sinh\xi\cos\vt\\
Q_2&=&\sinh\eta\sin\vt=\sin\mu\sinh\xi\sin\vt\\
Q_0&=&\sqrt{\sin^2\mu+\sinh^2\eta}=\sin\mu\cosh\xi
\ea
A boost is given by $\xi\ra\xi+\eps\xi_0$ and a rotation by
$\vt\ra\vt+\eps\vt_0$.
While the boosttransformations mix energy  and momentum space the rotations
leave the
momentum space hyperboloid invariant. They are symmetry operations on the
surface
and generated by:
\be
L=i\hbar(Q_2\frac{\dd}{\dd Q_1}-Q_1\frac{\dd}{\dd Q_2})\label{angul}
\ee
Although we did not quantize anything yet, we already added the $\hbar$ in the
definition of the operator to make it suitable for the quantum theory. Usually
we have two additional translational invariances in momentum space and one in
the direction of the energy, generated by ($\frac{\dd}{\dd p_0},\frac{\dd}{\dd
p_1},\frac{\dd}{\dd p_2}$) These generators will then be identified with the
coordinates $t,x,y$. This is however no longer true due to the fact that
momentum space is a hyperboloid. The operators $i\hbar\frac{\dd}{\dd Q_i}$
would simply move us out of the energy momentum surface defined by (\ref{hyp}).
Because we chose energy momentum space as a direct product of energy  and
momentum space, the definition of a time operator (for an off shell particle)
is simple:
\be
T=i\hbar\frac{\dd}{\dd H}
\ee
There are however two more operators that leave momentum space invariant
besides $L$. We will call them $X,Y$ for reasons that become clear later:
\footnote{In this paper we take $8\pi G=1$ unless it is stated differently}
\ba
X&=&i\hbar(Q_3\frac{\dd}{\dd Q_1}+Q_1\frac{\dd}{\dd Q_3})\label{X}\\
Y&=&i\hbar(Q_3\frac{\dd}{\dd Q_2}+Q_2\frac{\dd}{\dd Q_3})\label{Y}
\ea
We want to argue that these operators are the most natural ones to represent
the coordinates $x$ and $y$. Strictly speaking we should define these operators
differently. From (\ref{can}) we have:
\ba
&q_1=q\cos\vt&~~~~~x=i\hbar\frac{\dd}{\dd q_1}\\
&q_2=q\sin\vt&~~~~~y=i\hbar\frac{\dd}{\dd q_2}
\ea
If we work this out a bit further we have:
\ba
x&=&i\hbar(\cos\vt\frac{\dd}{\dd q}-\frac{\sin\vt}{q}\frac{\dd}{\dd\vt})\\
y&=&i\hbar(\sin\vt\frac{\dd}{\dd q}+\frac{\cos\vt}{q}\frac{\dd}{\dd\vt})
\ea
With respect to the Lorentz invariant measure on the hyperboloid:
\be
d\mu=\sinh\eta~d\eta~d\vt~~~~~q=2\eta
\ee
these operators are not Hermitian (due to the free hyperbolic angle $q$ in
these formulas). It makes more sense to break away from the usual quantization
and define $x$ and $y$ as the generators of the symmetry transformations on the
hyperboloid
(\ref{X},\ref{Y}) which are Hermitian with respect to this measure. Moreover we
think that we have a certain freedom to do this as long as in the low energy
limit things converge to the well known quantization schemes. What does
actually happen in the low energy limit? Clearly we want that $H,\eta\ll 1$.
This implies that we are close to the bottom of the hyperboloid where
everything seems flat (see figure (\ref{hyperboloid})). Approximately we have:
\ba
Q_0&\simeq& \tilde{H}\\
Q_1&\simeq&\eta\cos\vt=\eta_1\\
Q_2&\simeq&\eta\sin\vt=\eta_2\\
Q_3&\simeq& 1
\ea
The mass shell becomes the K.G. equation as we have seen in section 1
(\ref{KG}). Moreover we have:
\ba
X&\simeq&i\hbar\frac{\dd}{\dd \eta_1}\\
Y&\simeq&i\hbar\frac{\dd}{\dd\eta_2}
\ea
So in the limit we have indeed the usual quantization rules.

Next we will show that the operators $X$ and $Y$ do transform properly under
rotations. For that purpose we will introduce the Planck length
$\ell_P=G~\hbar$ (in units where $c=1$). It implies that the operators $X,Y$
become proportional to $\ell_P$ (i.e. $\hbar\ra\ell_P$ in (\ref{X},\ref{Y})).
We find:
\ba
{[L,X]}&=&i\hbar Y\\
{[L,Y]}&=&-i\hbar X\\
\ea
The commutator between the coordinates $X$ and $Y$ is quite peculiar and
explains the term `non commuting coordinates' in the title:
\be
{[X,Y]}=-i\frac{\ell^2_P}{\hbar}~L
\ee
In the large distance limit (low energy limit) $\ell_P\ll 1$ so that the term
proportional to $\ell_P^2$ is negligible. We find thus that the commutator
vanishes there. The non commutativity of
the coordinates may be regarded as a result of the peculiar definition of the
coordinates that
we made. However on very small distances it is not clear if a distance has a
precise meaning at all. It is more important that we have proper variables to
parametrize our phase space. The interpretation in terms of geometry has
probably only meaning for large distances as compared  with $\ell_P$. For this
same reason it does not matter what is the precise form of the momenta $P_1,
P_2$ as long as we have the right low energy limit. Whatever we choose we will
never find precisely the same commutators as in the `classical' case.
The simplest choice is:
\be
P_i=\frac{\hbar}{\ell_P}Q_i~~~~~i=1,2
\ee
We then find:
\ba
{[P_i,P_j]}&=&0\\
{[L,P_1]}&=&i\hbar P_2\\
{[L,P_2]}&=&-i\hbar P_1\\
{[X,P_1]}&=&i\hbar Q_3=i\sqrt{\hbar^2+(P_1~\ell_P)^2+(P_2~\ell_P)^2}
\simeq i\hbar(1+{\cal O}(\ell_P^2))\\
{[X,P_2]}&=&0\\
{[Y,P_1]}&=&0\\
{[Y,P_2]}&=&i\hbar Q_3=i\sqrt{\hbar^2+(P_1~\ell_P)^2+(P_2~\ell_P)^2}
\simeq i\hbar(1+{\cal O}(\ell_P^2))
\ea

\section{Two particle wavefunctions}

Now that we have esthablished what our momentum space looks like and what our
`quantum coordinates' are we will investigate the Hilbert space. Because we
have
non commuting coordinates in configuration space it proves easier to work in
momentum space. What are the most natural Hermitian operators to consider on
the hyperboloid? The analogon with the sphere might help here. There one takes
the Laplacian $\vec{L}^2$ on the sphere together with the angular momentum
$L_z$. This is precisely the situation that 't Hooft considered in
\cite{Hooft2}. These operators form a complete set of operators on the sphere.
We will do the same thing here and consider the Laplace (-Beltrami) operator,
$\Delta$, on the hyperboloid together with the angular momentum $L_z$:
\ba
\Delta &=&
-\hbar^2(\frac{1}{\sinh\eta}\frac{\dd}{\dd\eta}\sinh\eta\frac{\dd}{\dd\eta}
+\frac{1}{\sinh^2\eta}\frac{\dd^2}{\dd\vt^2})\label{Laplace}\\
&=&(\lambda^2+\frac{1}{4})\hbar^2~~~~~~~~~~~~~~~~~~~\lambda\in [0,\infty)
\ea
The invariant measure on the hyperboloid is
\be
d\mu=\sinh\eta~d\eta~d\vt
\ee
With respect to this measure we can write:
\be
\lambda^2+\frac{1}{4}=<\psi|\Delta\psi>=<O\psi|O\psi>=\parallel
O\psi\parallel^2
\ee
where:
\be
O=i\frac{\dd}{\dd\eta}+\frac{1}{\sinh\eta}\frac{\dd}{\dd\vt}
\ee
So we can write $\Delta=O^\dag O$ which proves that $\Delta$ is Hermitian and
positive definite. Let us introduce the angular functions:
\be
\phi_m(\vt)=\frac{1}{\sqrt{2\pi}}e^{im\vt}
\ee
They are of course the eigenfunctions of the Hermitian operator for the angular
momentum:
\be
L=-i\hbar\frac{\dd}{\dd\vt}\Rightarrow L\phi_m(\vt)=m\hbar\phi_m(\vt)
\ee
This is the same operator as (\ref{angul}) written in polar coordinates. The
possible (discrete) values of $m$ depend on the boundary condition on $\phi$.
First we will consider the (wrong) boundary condition:
$\psi(\vt+2\pi)=\psi(\vt)$ which gives integer valued values for $m$,
i.e. we neglect the cut out, wedgelike region in order to study the
influence
of the curved momentum space seperately. Later we will consider the right
boundary condition (\ref{boundarycondition}) leading to fractional angular
momentum.

We can seperate variables in the usual way:
\be
{\cal M}^\la_m(\eta,\vt)=F^\la_m(\eta)~\phi_m(\vt)
\ee
So that (\ref{Laplace}) becomes:
\be
 -\hbar^2(\frac{1}{\sinh\eta}\frac{\dd}{\dd\eta}\sinh\eta\frac{\dd}{\dd\eta}
-\frac{m^2}{\sinh^2\eta})F^\la_m(\eta)
=(\lambda^2+\frac{1}{4})\hbar^2~F^\la_m(\eta)~~~~~~\lambda\in [0,\infty)
\label{Laplace2}
\ee
The eigenvalues can be found by the following substitution:
\ba
&&\tanh^2\eta=z\label{sub1}\\
&&\sinh^2\eta=\frac{z}{1-z}\nn\\
&&\cosh^2\eta=\frac{1}{1-z}\nn
\ea
and
\be
F^\la_m(z)=z^{\frac{m}{2}}~(1-z)^{\frac{1}{2}(\frac{1}{2}-i\la)}~V^\la_m(z)
\label{sub2}
\ee
The Laplace equation for $F^\la_m(z)$ is then mapped to a hypergeometric
equation for $V^\la_m(z)$ from which we can  read off the solution (see
\cite{Limic}):
\ba
&&F^\la_m(\eta)=\mid \frac{\Gamma[\frac{1}{2}(|m|+i\la+\frac{1}{2})]
\Gamma[\frac{1}{2}(|m|+i\la+\frac{3}{2})]}{\sqrt{2\pi}\Gamma[i\la]
\Gamma[|m|+1]}
\mid \label{solution}\\
&& \tanh\eta^{|m|}\cosh\eta^{(i\la-\frac{1}{2})}~~
F[\frac{1}{2}(|m|-i\la+\frac{1}{2}), \frac{1}{2}(|m|-i\la+\frac{3}{2}),
|m|+1; \tanh^2\eta]\nn
\ea
Altogether we have a complete orthonormal basis for our Hilbert space:
$\{ {\cal M}^\la_m(\eta,\vt)=F^\la_m(\eta)~\phi_m(\vt)\}$:
\ba
&&\int_{0}^{2\pi}\int_0^\infty
\sinh\eta~d\eta~d\vt~\Ml~\Mrs=\de(\la-\la')\de_{m,m'}\\
&& \sum_{m=-\infty}^{\infty}\int_0^\infty
d\la~\Ml~\Mrr=\de(\cosh\eta-\cosh\eta')\de(\vt-\vt')
\ea
We can use this basis to define a Fourier transform from momentum space to
configuration space and back:
\ba
\hat{\psi}(\la,m)&=&\int_0^\infty\int_0^{2\pi}~\sinh\eta~d\eta~d\vt~
\psi(\eta,\vt)~\Ml\\
\psi(\eta,\vt)&=&\sum_{m=-\infty}^{\infty}\int_0^\infty d\la~\hat{\psi}(\la,m)
{}~\Mr
\ea
The coordinates that parametrize configuration space are the angular momentum
$m$ and the spectral parameter $\la$. But what does this $\la$ mean physically?
{}From section two we have:
\be
\frac{\Delta}{\hbar^2}=-\frac{L_z^2}{\hbar^2}+\frac{X^2}{\ell_P^2}
+\frac{Y^2}{\ell_P^2}=\la^2+\frac{1}{4}
\ee
Defining the operator for the distance between the particles: $R^2=X^2+Y^2$ we
find:
\be
R^2=(\frac{\Delta}{\hbar^2}+\frac{L_z^2}{\hbar^2})\ell_P^2
=(m^2+\la^2+\frac{1}{4})\ell_P^2 \geq 0
\ee
Notice that for the classical limit we have
\be
R\sim \la\ell_P~~~~~\la^2>>(m^2+\frac{1}{4})
\ee
In the classical limit (low energy limit) we therefore find that $\la$
represents the distance between the particles. Notice furthermore that there is
a smallest distance in our theory:
\be
R\geq R_{\bf min}=\frac{1}{2}\ell_P
\ee
This implies that the particles can never get closer to one another than
$\frac{1}{2}\ell_P$. This number thus serves as a natural cutoff in our theory.

\begin{figure}[t]
\centerline{\psfig{figure=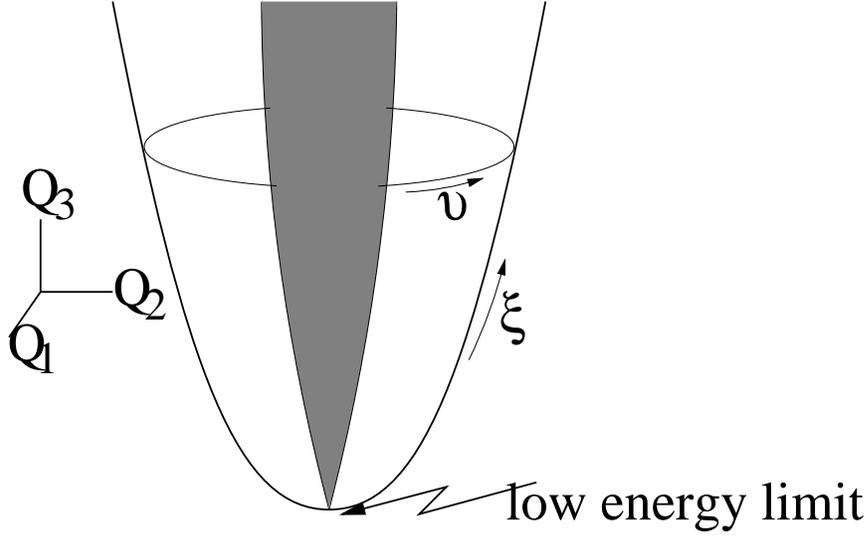,angle=-90,height=7cm}}
\caption{Momentum space represented as a hyperboloid with a wedgelike region
cut out.}
\label{hyperboloid}
\end{figure}

Let us return to the point where we had to choose a boundary condition for the
function $\phi_m(\vt)$. The correct boundary condition is of course:
\be
\psi(\vt+2\pi)=e^{imE}\psi(\vt)
\ee
giving rise to the quantization condition:
\be
m=\frac{2\pi~n}{2\pi-E}\equiv \frac{n}{\al}~~~n\in Z
\ee
This gives for $\phi_m(\vt)$:
\be
\phi_m(\vt)=\frac{1}{\sqrt{2\pi}}e^{i\frac{n}{\al}\vt}
\ee
Let us consider an effective particle\footnote{What we mean with an
effective particle is explained in the text below equation \ref{energy2}.}
on the mass shell and with a definite
direction $\vt$.
The solution in momentum space is:
\ba
\psi_E(\eta,\vt)&=&\de(\cosh\eta-\cosh\eta_0)\de(\vt-\vt_0)
\label{partonshell}\\
\cosh\eta_0&=&\frac{\cos\mu}{\cos(\frac{E}{4})}
\ea
What does this solution look like in our configuration space? Because we
consider fixed energy the value of $\al$ is fixed. In order to find a complete
set of eigenfunctions on this energy surface we solve the Laplace Beltrami
equation on hyperboloid with a wedgelike region cut out (see figure
(\ref{hyperboloid})).

So we have to solve again (\ref{Laplace2}) with $m=\frac{n}{\al}$. This is done
by the same substitutions (\ref{sub1},\ref{sub2}) and the solution is
again (\ref{solution}) but with $m$ replaced by $\frac{n}{\al}$. This is again
a complete set of eigenfunctions\footnote{It is easily shown that equation
(\ref{Laplace2}) is of the Schrodinger (or Sturm Liouville) type by the
following substitution:
$\phi=\sqrt{\sinh\eta}~F$. The equation then becomes:
\be
-\frac{\dd^2}{\dd\eta^2}\phi+\{\frac{\frac{1}{4}
\cosh^2\eta+m^2-\frac{1}{2}}{\sinh^2\eta}\}
\phi=(\la^2+\frac{1}{4})\phi
\ee
This equation has a complete set of eigenfunctions given by the hypergeometric
solutions
(\ref{solution}) with $m=\frac{n}{\al}$. In the limit where $\eta\gg 1$ we
have, by using the transformation formula's for the hypergeometric function,
the following asymptotic behaviour:
\be
F^\la_{\frac{n}{\al}}(\eta)\simeq
e^{-\frac{1}{2}\eta}\{A(\la,\frac{n}{\al})e^{i\la\eta}
+B(\la,\frac{n}{\al})e^{-i\la\eta}\}
\label{asymptoot}
\ee
Keeping in mind the measure $\sinh\eta~d\eta$ these functions are normalizable
as wave packets. We also see from this that the spectrum of $\Delta$ really
starts at $\frac{1}{4}$ because for for smaller values $\la$ becomes imaginary
and one of the terms in (\ref{asymptoot}) blows up.}
and we will use them to convert solution
(\ref{partonshell}) to the configuration space:
\ba
\psi_E(\la,n)&=&\int_1^\infty\int_0^{2\pi}
d\cosh\eta~d\vt~\de(\cosh\eta-\cosh\eta_0)~\de(\vt-\vt_0)
{}~F^\la_{\frac{n}{\al}}(\eta)e^{i\frac{n}{\al}\vt}\\
&=&F^\la_{\frac{n}{\al}}(\eta_0)e^{i\frac{n}{\al}\vt_0}
\ea
Notice that these functions are the analogon of the fractional Besselfunctions
$J_{\frac{n}{\al}}(\kappa r)e^{i\frac{n}{\al}\vt}$
of \cite{DesJac}. What if we consider an effective particle with different
energy? For this particle the deficit angle on the hyperboloid is different.
This means that we cannot use the same set of eigenfunctions as was used
before. Following the same calculation we have for this particle in
configuration space:
\be
\psi_{E'}(\la,n)=F^\la_{\frac{n}{\al'}}(\eta'_0)e^{i\frac{n}{\al'}\vt_0}
\ee
And thus the disaster happens that $\psi_E$ and $\psi_{E'}$ are not orthogonal.
We cannot construct a superposition of wavefunctions with different energy so
we cannot construct wavepackets in this theory. This situation is of course no
different than in the case where one used flat momentum space and found the
same problem for the fractional Besselfunctions. We think however that this
problem is not stressed sufficiently in the literature. We are not able to
solve this
problem now but are convinced that a in full theory of 2+1 quantum gravity
this issue must be
adressed. What we can do with the present understanding is calculate
scattering amplitudes, but ironically this will be the same in the
case of curved and flat momentum space. Because we fixed the the value of the
momentum $\eta=\eta_0$, the distance $R$ between the two particles is
completely unknown by the Heisenberg uncertainly relations. In other words,
because we do not know how to superpose wavefunctions with different energy it
is impossible to construct localized wavepackets and thus to conlude anything
about the spectrum of the $R$ operator. We expect however that the prediction
of a smallest distance will survive in the full theory as this is connected
with the curvature in momentum space and not with the problems mentioned above.
\section{Discussion}

In this paper we found that the conjugate momentum to the relative distance
between two particles in 2+1 dimensional gravity is a hyperbolic angle. This
caused us to consider quantization on a hyperbolic momentum space. The result
of this was that the model has a shortest distance built in which serves as a
natural cutoff in the theory. The fact that curvature in momentum space can
regularize the theory was already suggested by Snyder in 1947 \cite{Snyder}.
The beautiful idea behind this is that the model can still be covariant under
the full group of Lorenztransformations. The price one has to pay is the
introduction of non commuting coordinates. Also translational invariance is
lost. In this paper we kept translational invariance in the timelike direction
but introduced non commuting coordinates $X$ and $Y$. Because the Hamiltonian
is an angle we have discrete time. The distance between the particles has a
continuous spectrum but starts at $R_{\bf min}=\frac{1}{2}\ell_P$.

The boundary conditions on the wavefunctions prevented us from writing down a
complete set of orthonormal basisfunctions in the Hilbert space. The problem is
that wavefunctions with different energy obey different boundary conditions and
are therefore not orthogonal. In future work we hope to adress this issue.

The hope is of course that the introduction of a curved momentum space can act
as a covariant cutoff in a more advanced field theory, ultimately resulting in
a finite theory for quantum gravity.

\section{Acknowledgements}
I would like to thank the following people for the many discussions and
suggestions: G. 't Hooft,
H.J. Matschull, E. Verlinde and E. van der Ban.
\appendix

\renewcommand{\thesection}{Appendix \Alph{section}:}
\section{Massless particles}
A {\em massive} particle at position $a$ is described by the following
identification rule over a wedge:
\be
x'=a+e^{i p^aJ_a}(x-a)
\ee
where:
\be
p^a=(M\cosh\xi,M\sinh\xi\cos\vt,M\sinh\xi\sin\vt)
\ee
In order to describe a massless particle we take the limit $M\ra 0$ and
$\xi\ra\infty$ in such a way that:
\be
p^a\ra 2(\sg,\sg\cos\vt,\sg\sin\vt)\equiv 2\sg^a\label{limit}
\ee
The meaning of the parameter $2\sg$ will become clear in a moment. Clearly we
have that $\sg^a$ is a lightlike vector: $\sg^a\sg_a=0$. Notice also that the
particle moves with the speed of light: $v=\tanh\xi=1$. The energy of this
particle is given by the deficit angle of the wedge that we have cut out.
{}From \cite{Hooft2} we have the following relations:
\ba
\tan\frac{H}{2}&=&\cosh\xi~\tan\frac{M}{2}\ra\sg\\
\sin\frac{H}{2}&=&\frac{\tanh\eta}{\tanh\xi}\ra\tanh\eta\\
\cos\frac{H}{2}&=&\frac{\cos\frac{M}{2}}{\cosh\eta}\ra \frac{1}{\cosh\eta}\\
\sinh\eta&=&\sin\frac{M}{2}~\sinh\xi\ra\sg
\ea

\begin{figure}[t]
\centerline{\psfig{figure=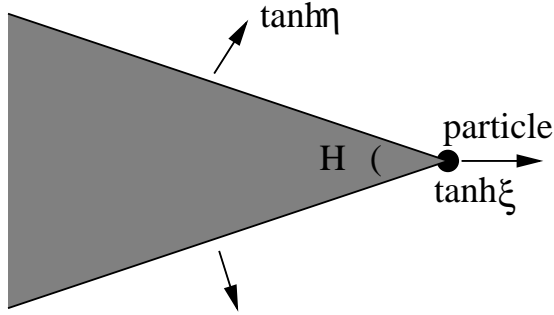,angle=-90,height=4cm}}
\caption{A `gravitating' particle in 2+1 dimensions deformes space time by
cutting
out a wedgelike region.}
\label{particle}
\end{figure}

So we must take $\sg=\sinh\eta$ where $\eta$ represents the rapidity with which
the boundary of the wedge moves (see figure (\ref{particle})). In the low
energy limit we have $H=2\eta$ where $2\eta$ is the momentum canonically
conjugate to the edgelength $L$, which in the limit fot small deficit angle is
just the distance from the origin to the particle. This is of course the
correct dispersion relation for a massless particle.

If we want to repeat the calculation of section one in the massless case we
must take the limit (\ref{limit}) in all formula's. The angle between the two
particles in their c.o.m. frame is then for instance:
\be
\cos\vfi=\frac{\sg^2-1}{\sg^2+1}
\ee
and the total energy becomes:
\be
\cos\frac{H}{4}=\frac{1}{\cosh\eta}=\frac{1}{\cosh\frac{q}{2}}
\ee
Notice that in the limit for small energy and
momentum we find the correct dispersion $H=2q$.

What does this massless limit imply for section two? Because we still want to
quantize on the hyperboloid, the definition of $Q_1,~Q_2$ and $Q_3$ does not
change. For $Q_0$ we now have:
\be
Q_0=\tan\tilde{H}
\ee
This definition indeed gives the massless Klein Gordon equation:
\be
Q_0^2-Q_1^2-Q_2^2=0
\ee
What we actually find is that we can safely take the limit $\mu\ra 0$ in all
the formula's.

\setcounter{equation}{0}
\renewcommand{\theequation}{\Alph{section}.\arabic{equation}}
\setcounter{figure}{0}
\renewcommand{\thefigure}{\Alph{section}.\arabic{figure}}
\section{The Dirac equation}
\setcounter{equation}{0}
In this appendix we will derive a Dirac equation that is linear in the
$Q's$. The derivation is valid for one particle in the polygon approach (with
variables $L,\eta$) and for two particles in their center of mass frame. The
starting point is the following set of equations:
\ba
-Q_0^2+Q_1^2+Q_2^2&=&-\sin^2\mu\\
-Q_3^2+Q_1^2+Q_2^2&=&-1\\
Q_3^2-Q_0^2&=&\cos^2\mu
\ea
We would like an equation that pushes time forward by one unit of time, i.e. we
are looking for an equation of the form:
\be
e^{i\tilde{H}}\psi=S\psi
\ee
where $S$ is an unitairy matrix in SU(2)\footnote{We take here
$\tilde{H}=\frac{H}{4}$ in the case of two particles in their c.o.m. frame and
$\tilde{H}=\frac{H_P}{2}$ in the case of one particle using polygon variables.
Accordingly we rescale time in both cases to $\tilde{t}=4t$ and
$\tilde{t}=2t_P$ respectively.}.
We know from (\ref{energy2}) that:
\ba
Q_3~\cos\tilde{H}&=&\cos\mu\Rightarrow\\
Q_3~\sin\tilde{H}&=&\sqrt{Q_3^2-\cos^2\mu}=Q_0\label{Q0}\\
&=&\sqrt{Q_1^2+Q_2^2+\sin^2\mu}
\ea
Combining these equations gives:
\be
Q_3~e^{i\tilde{H}}\psi=(\cos\mu+i\sqrt{Q_1^2+Q_2^2+\sin^2\mu})\psi
\ee
Using Dirac's trick to linearize this equation gives:
\be
Q_3~e^{i\tilde{H}}\psi=(\cos\mu+i\bt\sin\mu+i\al_1Q_1+i\al_2Q_2)\psi
\ee
Here we can take the Dirac matrices to be the Pauli matrices: $\bt=\sg_3$ and
$\al_i=\sg_i$. In matrix notation:
\be
e^{i\tilde{H}}\psi=\frac{1}{\cosh\eta}\left( \begin{array}{cc}
e^{i\mu} & i\sinh\eta~e^{-i\vt}\\
i\sinh\eta~e^{i\vt} & e^{-i\mu} \end{array}\right)\psi
\ee
Clearly the right hand side is an element of SU(2) which assures conservation
of probability. It is also clear that this equation contains particles and
anti particles as a solution which will become part of our model if we use this
equation.

What about covariance? Using (\ref{Q0}) we can write:
\be
(Q_3~\cos\tilde{H}+iQ_0)\psi=\cos\mu~\psi+i(\sg_1Q_1+\sg_2Q_2+\sg_3\sin\mu)\psi
\ee
We can rewrite this as:
\be
\sg_3(Q_3~\cos\tilde{H}-\cos\mu)\psi-i(\ga^aQ_a+\sin\mu)\psi=0
\ee
where we have defined: $\ga^0=-\sg_3$ and $\ga^i=\sg_3\sg_i$. If $\psi$
transforms according to the two dimensional representation of the Lorentzgroup
SU(1,1) and ($Q_0,Q_1,Q_2$) transforms as a Lorentzvector under SO(2,1) we can
derive with the usual argument\footnote{The usual argument being that
$h~\ga^a~h^{-1}=L^a_{~b}\ga^b$ where $L^a_{~b}\in$ SO(2,1) and $h\in$ SU(1,1).}
that the second term is covariant. However, the first term does not transform
covariantly as we have:
\be
Q_3=\sqrt{\cos^2\mu+\sin^2\mu~\cosh^2\xi}
\ee
and a boost is generated by $\xi\ra\xi+\eps\xi_0$. We must conclude that the
Dirac equation is only Lorentz invariant on the mass shell:
$Q_3~\cos\tilde{H}=\cos\mu$.

Another important point is locality of the equation in configuration space. The
matrix on the right hand side has by no means a `local action' on the states
$F^\la_m(\eta)\phi_m(\vt)$. So it is not clear if this equation is an
improvement over the Klein Gordon equation in that respect.

Still another difficulty is the implementation of the boundary condition. If we
fix the energy as in section 3 we have to consider this equation on a
hyperboloid with a wedge cut out. The action of $Q_3$ is still well defined but
what about $Q_1$ and $Q_2$? They contain $\cos\vt$ and $\sin\vt$ while the
basis is built from the angular functions:
$\phi_{\frac{n}{\al}}(\vt)=\frac{1}{\sqrt{2\pi}}e^{i\frac{n}{\al}\vt}$. The
action of the operators $Q_\pm\equiv Q_1\pm iQ_2$ on these states is:
\be 
Q_\pm~\phi_{\frac{n}{\al}}(\vt)=\sinh\eta\frac{1}{\sqrt{2\pi}}
e^{i(\frac{n}{\al}\pm 1)\vt}
\ee
These states are however no longer in the basis set, so that $Q_\pm$ throws
them out of the Hilbert space. Clearly if we want to take the boundary
condition
into account correctly we need to modify this Dirac equation. For one particle
described by polygon variables one can view this Dirac equation as an analogon
of the Dirac equation derived by 't Hooft in \cite{Hooft3}.

\end{document}